\documentclass[preprint,12pt]{elsarticle}

\usepackage{amssymb}
\usepackage[T1]{fontenc}
\usepackage[latin9]{inputenc}   
\usepackage{color}
\usepackage{textcomp}
\usepackage{amstext}
\usepackage{graphicx}
\usepackage{epstopdf}

\journal{Journal of Non-Crystalline Solids}

\begin{document}

\begin{frontmatter}

\title{Dynamic Light Scattering Study of Temperature and pH Sensitive Colloidal Microgels}
%\maketitle

%% use optional labels to link authors explicitly to addresses:
%% \author[label1,label2]{}
%% \address[label1]{}
%% \address[label2]{}

\author[Roma3]{Valentina Nigro\footnote{Corresponding author: valentina.nigro@gmail.com}}
\author[IPCF-RM,DIP]{Roberta Angelini}
\author[IPCF-PI]{Monica Bertoldo}
\author[PI]{Valter Castelvetro}
\author[DIP,IIT]{Giancarlo Ruocco}
\author[IPCF-RM,DIP]{Barbara Ruzicka}

\address[Roma3] {Dipartimento di Fisica, Universit\`a degli Studi  di Roma Tre, Via della Vasca Navale 84, 00146 Roma, Italy.}
\address[IPCF-RM]{Istituto per i Processi Chimico-Fisici del Consiglio Nazionale delle Ricerche (IPCF-CNR), UOS Roma, Pz.le Aldo Moro 5, I-00185 Roma, Italy.}
\address[DIP]{Dipartimento di Fisica, Sapienza Universit$\grave{a}$ di Roma, Pz.le Aldo Moro 5, I-00185, Italy.}
\address[IPCF-PI]{Istituto per i Processi Chimico-Fisici del Consiglio Nazionale delle Ricerche (IPCF-CNR), Area della Ricerca, Via G.Moruzzi 1, I-56124 Pisa, Italy.}
\address[PI]{Dipartimento di Chimica e Chimica Industriale, Universit$\grave{a}$ di Pisa, via Risorgimento 35, I-56126 Pisa, Italy.}
\address[IIT]{Center for Life Nano Science, IIT@Sapienza, Istituto Italiano di Tecnologia, Viale Regina Elena 291, 00161 Roma, Italy.}

\begin{abstract}
Microgel particles composed of Interpenetrated Polymer Networks (IPN) of
poly(N-isopropylacrylamide) (PNIPAM) and poly(acrylic acid) (PAAc)
dispersed in water have been investigated through dynamic light
scattering. The study of the temperature, concentration and pH
dependence of the relaxation time has highlighted the existence of
a thermoreversible transition corresponding to the
swollen-shrunken volume phase transition. The presence of PAAc
introduces an additional pH-sensitivity with respect to the
temperature-sensitivity due to PNIPAM and leads to interesting
differences in the transition process at acid and neutral pH.
\end{abstract}


\end{frontmatter}

\section{Introduction}
\label{Intro}

Colloidal systems have long been the subject of intense research
either for theoretical implications and for technological
applications. They are very good model systems for understanding
the general problem of dynamic arrest, due to their larger
tunability with respect to atomic and molecular
glasses~\cite{SciAdvPhys2005,TrappeCOCIS2004,PoonCOCIS1998,ZacJPCM2008}.
Moreover, thanks to their dimensions, colloidal systems can be
easily investigated through conventional techniques, such as
dynamic light scattering (DLS) and optical microscopy. The control
of their interparticle potential~\cite{LikosPhysRep2001} by tuning
external parameters such as packing fraction, waiting time and
ionic strength, has given rise to exotic phase diagrams with
different arrested states (such as gels~\cite{LuNat2008,
RoyallNatMat2008, RuzickaNatMat2011} and
glasses~\cite{PuseyNat1986, ImhofPRL1995}) and unusual glass-glass
transitions~\cite{PhamScience2002, EckertPRL2002, AngeliniNC2014}.
At variance with hard colloids soft colloids are characterized by
an interparticle potential with a finite repulsion at or beyond
contact. Theoretical
studies~\cite{LikosJPCM2002,RamirezJPCM2009,HeyesJCP2009} have
indicated the existence of an even more complex phase behavior not
experimentally reproduced up to now. Among  soft colloidal systems
 microgels, aqueous dispersions of nanometre- or
micrometre-sized hydrogel particles,  allow furthermore to
modulate the interaction potential through temperature and/or pH,
unlike in ordinary colloids.  Recent experimental studies have
shown the evidence of unusual transitions between different
arrested states~\cite{WangChemPhys2014} and the possibility to
make strong glasses~\cite{MattssonNature2009}. Moreover microgels
have been largely studied in the last years because of their
versatility and high sensitivity to external stimuli such as pH,
temperature, electric field, ionic strength, solvent, external
stress or light and are therefore particularly attractive smart
materials~\cite{SaundersACIS1999, PeltonAdvColloid2000,
VinogradovCurrPharmDes2006, DasAnnRevMR2006, KargCOCIS2009}.
Furthermore they find many applications in a lot of different
fields such as in agriculture, construction, cosmetic and
pharmaceutics industries, in artificial organs and tissue
engineering~\cite{PeltonAdvColloid2000, DasAnnRevMR2006,
ParkBiomat2013, BajpaiProgPolym2008, HamidiDrugDeliv2008,
SchexnailderCollPolym2009, SmeetsPolymSci2013, SuBiomacro2008}.

One of the most studied responsive microgel is based on the
 poly(N-isopropylacrylamide) also known as PNIPAM, a
 thermo-sensitive polymer. PNIPAM microgels were investigated for the first time in 1986 by
Robert Pelton and Philip Chibante~\cite{PeltonColloids1986}, since
then they have been widely studied both experimentally and
theoretically and a clear picture of preparation, characterization
and applications has been provided~\cite{SaundersACIS1999,
PeltonAdvColloid2000, DasAnnRevMR2006, KargCOCIS2009,
LuProgPolSci2011}. PNIPAM microgels responsiveness is strongly
dependent on the thermo-sensitivity of PNIPAM that presents a
Lower Critical Solution Temperature in water at about 305 K.  At
room temperature indeed, it is found in a swollen state, the
polymer is hydrophilic and a great amount of water is retained; by
increasing temperature above 308-311 K the polymer becomes
hydrophobic, water is completely expelled and a shrunken state is
found giving rise to a volume phase transition (VPT). As a result,
any microgel based on PNIPAM undergoes the characteristic VPT
due to the temperature sensitivity of PNIPAM~\cite{WuMacromol2003}. Phase diagram
~\cite{WuPRL2003, PaloliSM2012, WangChemPhys2014} and important
details on the gel structure of PNIPAM microgels near the volume
phase transition have been obtained~\cite{GaoLangmuir2002,
TangLangmuir2004}. It has been recently shown that the microgel
swelling/deswelling behavior can be strongly affected by
concentration~\cite{TanPolymers2010,WangChemPhys2014}, by
 solvents~\cite{ZhuMacroChemPhys1999} and by
synthesis procedure such as growing number of cross-linking
points~\cite{KratzBerBunsenges11998, KratzPolymer2001}, different
reaction pH conditions~\cite{BaoMacromol2006} or by introducing
additives into the PNIPAM network~\cite{HellwegLangmuir2004}.

In this context PNIPAM microgels containing another specie as
co\textendash{}monomer or interpenetrated polymer result to be
even more interesting systems since they retain the main
properties of the constituent polymer. In particular, adding
acrilic acid (AAc) to PNIPAM microgel provides a pH-sensitivity to
the system that leads to a more complex phase behavior. In this
way a thermo and pH-sensitive system is obtained. As in the case
of pure PNIPAM microgel a volume phase transition with temperature
is observed, although with a remarkable reduction in the swelling
capability as the AAc concentration is
increased~\cite{HuAdvMater2004,MaColloidInt2010}. Indeed the
volume phase transition of these microgels is strongly dependent
on the effective charge density controlled by the content of AAc
monomer~\cite{HuAdvMater2004, MaColloidInt2010}, on the pH of the
suspension~\cite{KratzColloids2000, KratzBerBunsenges21998,
XiaLangmuir2004, JonesMacromol2000} and on the salt
concentration~\cite{KratzColloids2000, XiongColloidSurf2011}. In
this framework the synthesis procedure plays an important role.
Indeed, AAc can be incorporated into PNIPAM either by random
copolymerization (PNIPAM-co-AAc)~\cite{KratzColloids2000,
KratzBerBunsenges21998, JonesMacromol2000, XiongColloidSurf2011,
MengPhysChem2007, LyonJPCB2004, HolmqvistPRL2012, DebordJPCB2003}
or by polymer interpenetration (IPN
PNIPAM-PAAc)~\cite{HuAdvMater2004, XiaLangmuir2004, XiaJCRel2005,
ZhouBio2008, XingCollPolym2010, LiuPolymers2012}. In the first
case particles are composed of a single network of both monomers,
with properties dependent on the monomer
ratio~\cite{KratzColloids2000}. Conversely, in the second case
microgels are made of two interpenetrated homopolymeric networks
of PNIPAM and PAAc, with the same independent response as the two
components to external stimuli~\cite{ZhouBio2008}. This important
characteristic of interpenetrating polymer network (IPN) microgels
makes the mutual interference between the temperature-responsive
and pH-responsive polymers largely reduced making the temperature
dependence of the VPT unchanged with respect to the case of pure
PNIPAM microgel (close to physiological temperature) and the IPN
microgel more suitable for applications in controlled drug release
and sensors~\cite{ParkBiomat2013, ChenEnvChem2013}. However
investigations on IPN PNIPAM-PAAc microgels are scarce and only a
preliminary temperature-concentration phase diagram has been
reported up to now~\cite{HuAdvMater2004}, leaving unexplored the
dependence on pH, ionic strength and crosslink density. A few
investigations through DLS, rheology
and microscopic techniques have confirmed the existence of the VPT
transition~\cite{XiaLangmuir2004,XiaJCRel2005, ZhouBio2008,
XingCollPolym2010, LiuPolymers2012}. Nevertheless the scenario is
far from being completely clear. Furthermore other open issues are
related to the possibility to vary the \textit{softness} of
PNIPAM-PAAc microgel particles that makes this system a suitable
and unique model to provide insights into the glass formation in
molecular systems ~\cite{MattssonNature2009}.

In this work, through DLS experiments, we study the dynamics of
aqueous dispersions of IPN microgels of PNIPAM and PAAc as a
function of temperature, pH, concentration and momentum transfer
$Q$ across the VPT transition. The relaxation time $\tau$  shows a
clear dependence on temperature and new features related to both
concentration and pH. Furthermore the investigation at different
lengthscales has highlighted a $\tau \propto Q^{-n}$ dependence
with $n>$2. These results open the way to further investigations
at different length and time scales.

\section{Material and Methods}
\label{Material and Methods}
  \subsection{Materials}
  \label{Materials}

    \paragraph{Materials}
Both N-isopropylacrylamide (NIPAM) from Sigma-Aldrich and
N,N'-methylene-bis-acrylamide (BIS) from Eastman Kodak were
purified by recrystallization from  hexane and
methanol, respectively, dried under reduced pressure (0.01 mmHg) at room
temperature and stored at 253 K until used. Acrylic acid (AAc)
from Sigma-Aldrich was purified by distillation (40 mmHg, 337 K)
under nitrogen atmosphere in the presence of hydroquinone and
stored at 253 K until used. Sodium dodecyl sulphate (SDS), purity
98 \%, potassium persulfate (KPS), purity 98 \%, ammonium
persulfate, purity 98 \%,
N,N,N\textasciiacute{},N\textasciiacute{}-tetramethylethylenediamine
(TEMED), purity 99 \%, ethylenediaminetetraacetic acid (EDTA),
NaHCO$_3$, were all purchased from Sigma-Aldrich and used as
received. Ultrapure water (resistivity: 18.2 M$\Omega$/cm at 298
K) was obtained with Sarium\textsuperscript{\textregistered} pro Ultrapure Water purification
Systems, Sartorius Stedim from demineralized water. All other
solvents were RP grade (Carlo Erba) and were used as received.
Before use dialysis tubing cellulose membrane, cut-off 14000 Da,
from Sigma-Aldrich, was washed in running distilled water for 3 h,
treated at 343 K for 10 min into a solution containing a  3.0 \%
weight concentration of NaHCO$_3$ and 0.4 \%  of EDTA, rinsed in
distilled water at 343 K for 10 min and finally in fresh distilled
water at room temperature for 2 h.

   \paragraph{Synthesis of IPN microgels}
The IPN microgels were synthesized by a sequential  free radical
polymerization method: in the first step PNIPAM micro-particles
were synthesized by precipitation polymerization and in the second
step acrylic acid was polymerized into the preformed PNIPAM
network~\cite{HuAdvMater2004}. (4.0850 $\pm$ 0.0001) g of NIPAM,
(0.0695 $\pm$ 0.0001) g of BIS and (0.5990 $\pm$ 0.0001) g of SDS
were solubilized in 300 mL of ultrapure water and transferred into
a 500 mL five-necked jacked reactor equipped with condenser and
mechanical stirrer. The solution was deoxygenated by bubbling
nitrogen inside for 30 min and then heated at (273.0 $\pm$ 0.3) K.
 (0.1780 $\pm$ 0.0001) g of KPS (dissolved in 5 mL of deoxygenated water) was added to initiate the polymerization and the reaction was allowed to proceed for 4 h.
The resultant PNIPAM microgel was purified by dialysis against
distilled water with frequent water change for 2 weeks. The final
weight concentration  and diameter of PNIPAM micro-particles were
1.02 \%  and 80 nm (at 298 K) as determined respectively by
gravimetric  and DLS measurements. (65.45 $\pm$ 0.01) g of the
recovered PNIPAM dispersion and (0.50 $\pm$ 0.01) g of BIS were
mixed and diluted with ultrapure water up to a volume of 320 mL.
The mixture was transferred into a 500 mL five-necked jacketed
reactor kept at (295 $\pm$ 1) K by circulating water and
deoxygenated by bubbling nitrogen inside for 1 h. 2.3 mL of AAc
and (0.2016 $\pm$ 0.0001) g of TEMED were added and the
polymerization was started with (0.2078 $\pm$ 0.0001) g of
ammonium persulfate (dissolved in 5 mL of deoxygenated water). The
reaction was allowed to proceed for 65 min and then stopped by
exposing to air. The obtained IPN microgel was purified by
dialysis against distilled water with frequent water change for 2
weeks, and then lyophilized to constant weight. The poly(acrylic
acid) content determined  by acid/base titration was 6.6 \% weight
concentration (PAAc/IPN).

\paragraph{Sample preparation}
IPN samples were prepared by dispersing lyophilized IPN into
ultrapure water at weight concentration 1.0 and 3.0 \% by magnetic
stirring for at least 3 h. Samples at different concentrations
were obtained by dilution.

\paragraph{Characterization}
The poly(acrylic acid) content in 10 g of IPN dispersion was
determined  by addition of 11 mL of 0.1 M NaOH followed by
potentiometric back titration with 0.1 M hydrochloric acid.
Concentration of dispersion was determined from the weight of the
residuum after water removal by lyophilisation, corrected for the
moisture residual amount obtained by thermogravimetric analysis
(TGA). This was accomplished with a SII Nano-Technology EXTAR
TG/DTA7220 thermal analyser at 275 K/min in nitrogen atmosphere
(200 mL/min). 5 mg of sample in an alumina pan was analysed in the
(313-473) K temperature range and the weight loss was assumed as
moisture content.

%DLS analysis was accomplished with a Brookhaven 90 Plus Dynamic
%Laser Light Scattering instrument collecting the scattered light
%at 90 °.

 \subsection{Experimental Methods}
 \label{Experimental Methods}
\begin{figure}
\begin{centering}
\includegraphics[width=8cm]{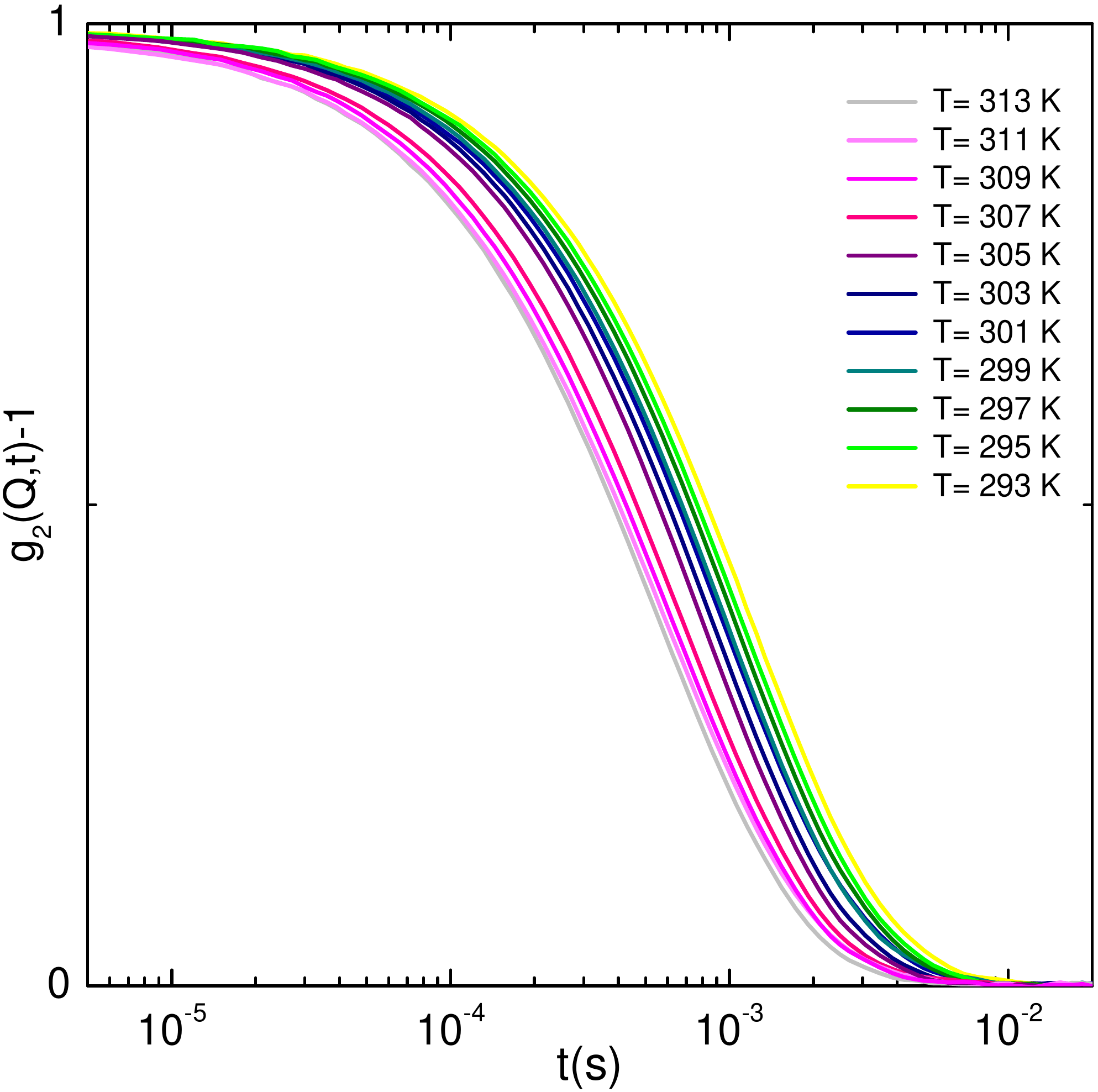}
\par\end{centering}

\caption{\label{fig:corr func}Normalized intensity autocorrelation
functions collected at $\theta$=90\textdegree of an IPN sample at $C_w=0.10$ \% and pH 7 for the indicated temperatures.}

\end{figure}

In Fig.~\ref{fig:corr func} the typical behavior of the normalized
intensity autocorrelation functions for an IPN sample at weight
concentration $C_w$=0.10 \%, pH 7 and $\theta$=90\textdegree for
the indicated temperatures is shown.
DLS measurements have been performed with a multiangle light
scattering setup. The monochromatic and polarized beam  emitted
from a solid state laser with a wavelength $\lambda$=642 nm and a
power of 100 mW is focused on the sample placed in a cylindrical
VAT for index matching and temperature control. The scattered
intensity is collected at five different values of the scattering
angle $\theta$=30\textdegree, 50\textdegree, 70\textdegree,
90\textdegree, 110\textdegree, that correspond to five different
values of the momentum transfer $Q$, according to the relation
Q=(4$\pi$n/$\lambda$) sin($\theta$/2). Single mode optical fibers
coupled to collimators collect the scattered light as a function
of time and scattering vector. In this way one can measure
simultaneously the normalized intensity autocorrelation function
$g_2(Q,t)=<I(Q,t)I(Q,0)>/<I(Q,0)>^{2}$ at five different Q values
with an high coherence factor close to the ideal unit value.
Measurements have been performed as a function of temperature
across the VPT.
As commonly known  the intensity correlation functions of most
colloidal systems cannot be well described  through a single
exponential decay but the Kohlrausch-Williams-Watts
expression~\cite{KohlrauschAnnPhys1854, WilliamsFaradayTrans1970} is generally used:
\begin{equation}
g_2(Q,t)=1+b[e^{-(t/\tau)^{\beta}}]^{2} \label{Eqfit}
\end{equation}
where $b$ is the coherence factor, $\tau$ is an ''effective''
relaxation time and $\beta$ generally
describes the deviation from the simple exponential decay ($\beta$
= 1). The different relaxation time present in glassy materials
lead to a stretching of the correlation functions (here referred
to as ''stretched behavior'') characterized by an exponent $\beta$
< 1 which can be related to the distribution of the relaxation
times. In the case of Brownian diffusion
it is possible to relate the relaxation time to the translational
diffusion coefficient $D_t$ through the relation
$\tau=1/(Q^2D_t)$. In the limit of non interacting spherical
particles  the Stokes Einstein relation $R=K_B T /6 \pi \eta D_t$
allows, known the viscosity $\eta$, to calculate the hydrodynamic
radius from the diffusion coefficient.
\begin{figure}
\begin{centering}
\includegraphics[width=8cm]{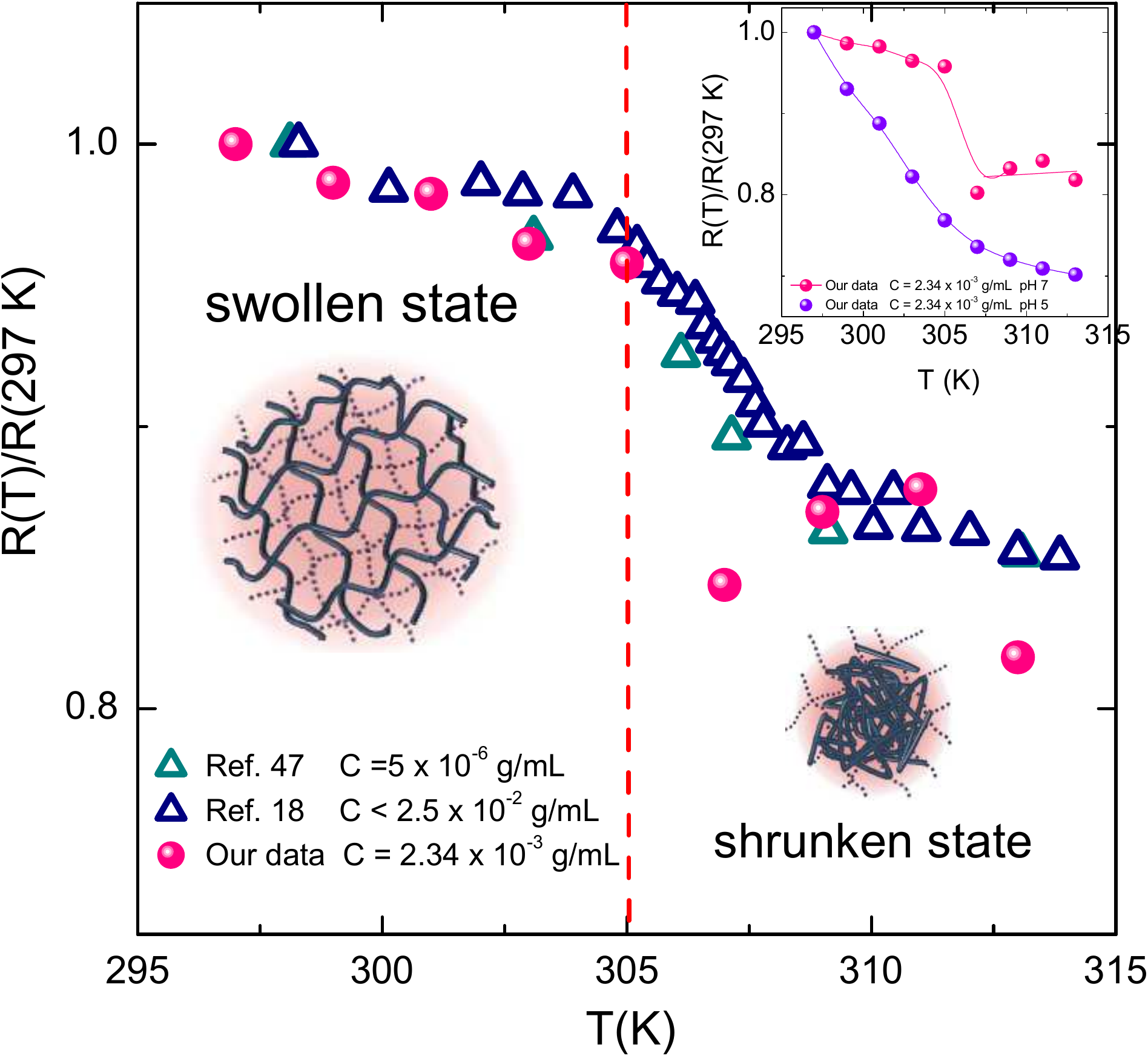}
\par\end{centering}

\caption{\label{Rnorm}Normalized radius as obtained from DLS
measurements for an IPN sample at $C_w=0.10$ \% (equivalent to a weight/volume concentration of $2.34\times10^{-3}$ g/mL), pH 7 and
$\theta$=90\textdegree compared with results from
ref.~\cite{XiaLangmuir2004, MattssonNature2009}. Inset: comparison between our results for the temperature dependence of the normalized radius at pH 7 and at pH 5. Full lines are guides to the eyes.}

\end{figure}

In Fig.~\ref{Rnorm}  the temperature dependence of $R$ at
pH 7 normalized with respect to the value $R$= (188.3$\pm$0.6)nm at
$T$=297 K is shown together with the results from previous works
at different concentrations~\cite{XiaLangmuir2004,
MattssonNature2009}. In the inset of Fig.~\ref{Rnorm} the
comparison between our results obtained for pH 7 and pH 5 (using
$R$=(135.2$\pm$0.5) nm at $T$=297 K) is also shown. The hydrodynamic radii of this work have been calculated by using the viscosity of the solvent. Our results are in agreement with previous
works~\cite{XiaLangmuir2004, MattssonNature2009} with a slight
discrepancy at high temperatures above the VPT. This can be
attributed to different reasons: the approximation of using the
solvent viscosity and the possible failure of the Stokes-Einstein
relation in the shrunken high temperature state due to non
spherical and/or interacting particles. For all these reasons in
the following we will always refer to the behavior of the
relaxation time rather than to the hydrodynamic radius.

\section{Results and Discussion}
\label{Results}

The DLS technique has been used to characterize the swelling
behavior of the IPN microgel in the temperature range 293 K$\leq
T\leq$ 313 K where the volume phase transition is expected to
occur. In order to neglect interparticle interactions and avoid
phase separation diluted solutions at four different weight
concentrations $C_w$=0.10 \%, $C_w$=0.15 \%, $C_w$=0.20 \%,
$C_w$=0.30 \% have been studied. To test reproducibility
measurements have been repeated several times. The pH dependence
of the suspensions has been investigated  both for acid and
neutral solution at pH 5 and 7, respectively. In this way a clear
picture of the microgel behavior as a function of temperature, pH
and concentration is drawn.
\begin{figure}
\begin{centering}
\includegraphics[width=8cm]{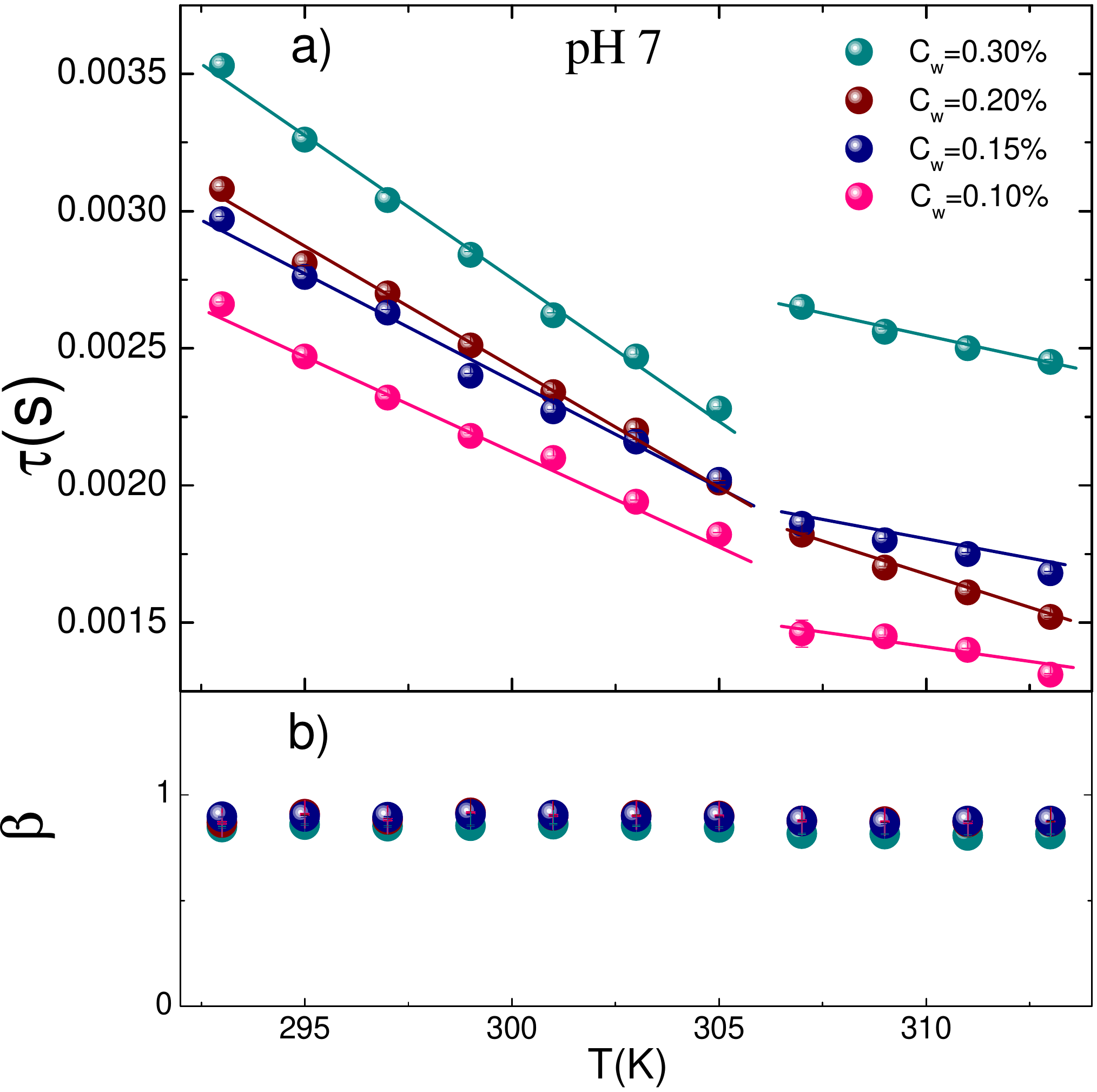}
\par\end{centering}
\caption{\label{fig:tau beta vs T_pH7} (a) Relaxation times and
(b) stretching parameter from Eq.~\ref{Eqfit} as a function of
temperature at pH 7 for the indicated concentrations. Full lines
in (a) are guides to the eyes.}
\end{figure}
In Fig.~\ref{fig:corr func} the intensity correlation functions
show a clear transition occurring above T=305 K. The volume phase transition from a swollen to a shrunken state is associated to a dynamical transition evidenced by looking at the relaxation time
obtained through a fit of the intensity correlation functions
according to Eq.~\ref{Eqfit}. Its behavior with temperature is
reported in Fig.~\ref{fig:tau beta vs T_pH7}: as temperature is
increased the relaxation time shows a slight decrease until the
transition is approached around 305 K. Thereafter, depending on
concentrations, different behaviors are observed. For the lowest
concentrated sample at $C_w$=0.10 \% after 305 K the relaxation
time abruptly reaches its lowest value, corresponding to a
transition of the microgel particles from the swollen to the
shrunken state. The aforementioned transition of IPN has been
reported to be smoother with respect to the case of pure PNIPAM
microgel due to the presence of the acrylic acid that makes the
swelling capability of the microgel greatly
reduced~\cite{HuAdvMater2004, XiaLangmuir2004, JonesMacromol2000}.
Our results indicate that at increasing concentration the jump
becomes smaller and smaller with an interesting swap in trend for
the highest concentrated sample at $C_w$=0.30 \%. Moreover, at
increasing concentration the relaxation time increases. On the
contrary the stretching parameter $\beta$ does not show any change
with temperature and concentration, remaining just below one,
indicating correlation functions slightly stretched.

\begin{figure}
\begin{centering}
\includegraphics[width=8cm]{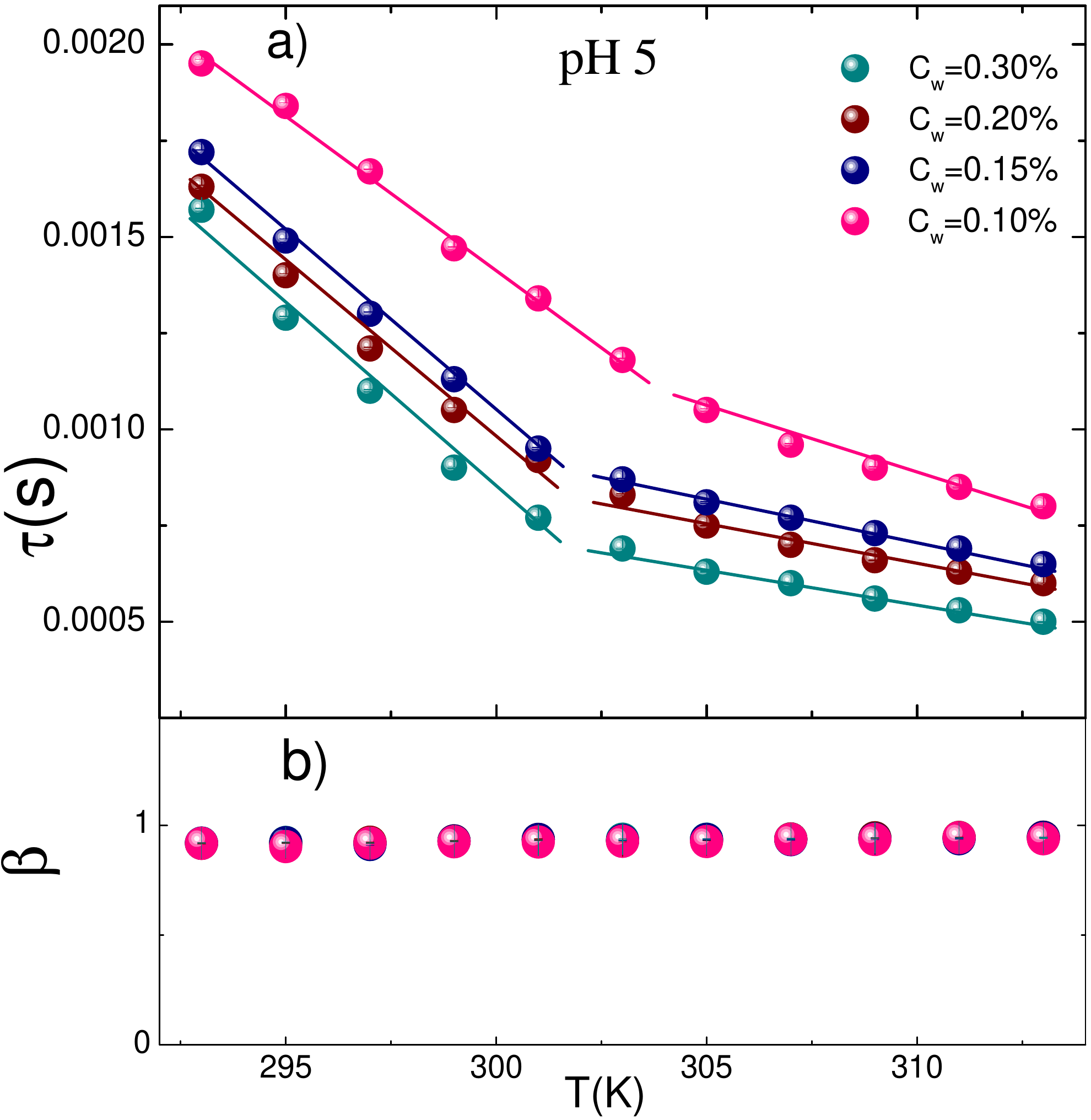}
\par\end{centering}
\caption{\label{fig:tau beta vs T_pH5} (a) Relaxation times and
(b) stretching parameter from Eq.~\ref{Eqfit} as a function of
temperature at pH 5 for the indicated concentrations. Full lines
in (a) are guides to the eyes.}
\end{figure}

This behavior is strongly affected by the pH of the solution as
shown in Fig.~\ref{fig:tau beta vs T_pH5} where relaxation time
and stretching parameters as a function of temperature under acid
conditions, at pH 5, are reported. At variance with neutral pH
solutions in this case a sharp transition is never observed and as
temperature is increased the relaxation time always decreases.
Furthermore an opposite trend with respect to concentration is
observed: with increasing concentration the relaxation time
decreases. The comparison between Fig.~\ref{fig:tau beta vs
T_pH7}a and Fig.~\ref{fig:tau beta vs T_pH5}a shows also that the
relaxation times are always smaller in acid conditions. As in the
case of neutral pH the stretching coefficient $\beta$ is neither
temperature nor concentration dependent and is always slightly
below 1.

Therefore the pH of the solution strongly affects the relaxation
times behavior: at neutral pH its values are higher and the
transition appears to be more evident than in the case at acid pH.
This dependence of the transition not only on concentration but
also on pH confirms that the effect observed is due to the
presence of PAAc. On the contrary the stretching parameter is
insensitive even to pH changes.
\begin{figure}
\begin{centering}
\includegraphics[width=8cm]{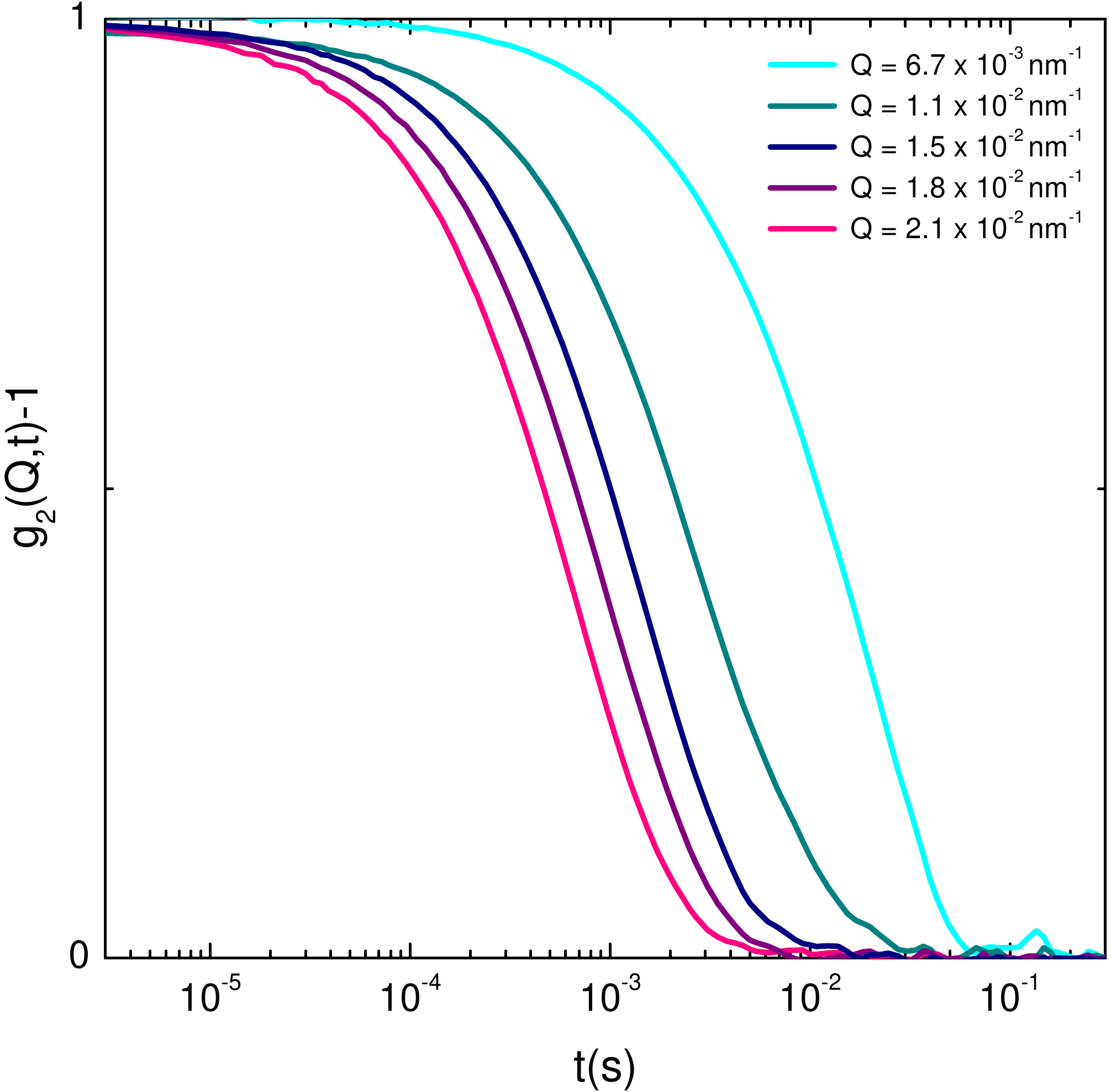}
\par\end{centering}
\caption{\label{fig:g(t)Q}Normalized intensity autocorrelation
curves at $C_w$=0.20 \%, T=303 K and pH 7 for the indicated values
of the exchanged momentum $Q$.}
\end{figure}

\begin{figure}
\begin{centering}
\includegraphics[width=8cm]{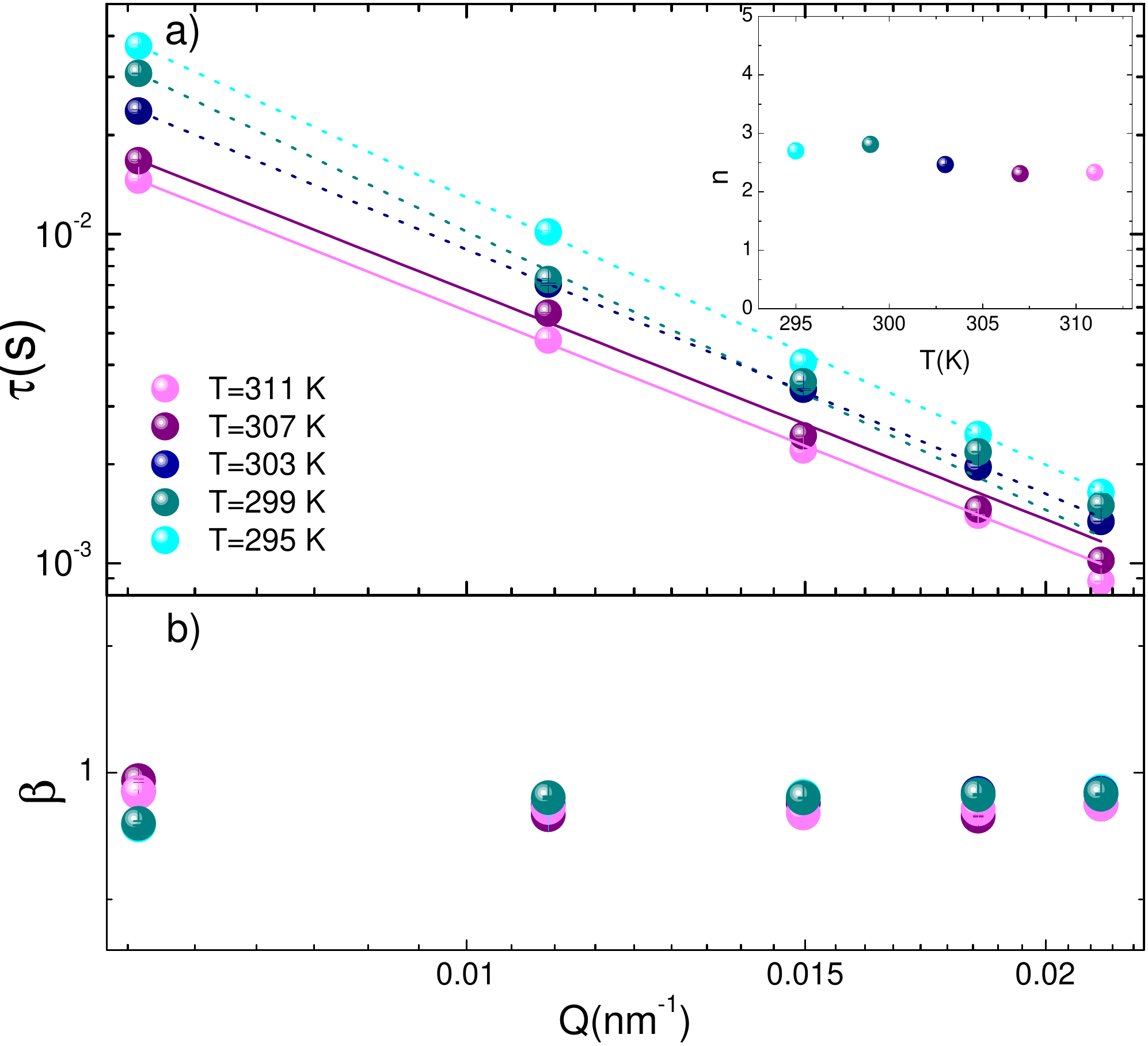}
\par\end{centering}

\caption{\label{fig:tau beta vs Q}(a) Relaxation time and (b)
stretching parameter from Eq.~\ref{Eqfit} as a function of
momentum transfer $Q$  at $C_w$=0.10 \%, pH 7 for the indicated
temperatures. Full lines are fits through Eq.~\ref{EqPowerLaw}
 with n=2.33 $\pm$ 0.03 (T=311 K), n=2.31 $\pm$ 0.07  (T=307 K), n=2.465 $\pm$ 0.008 (T=303 K), n=2.81 $\pm$ 0.08 (T=299 K), n=2.70 $\pm$ 0.03 (T=295 K). Inset: behavior of the exponent n as a function of temperature at $C_w$=0.10 \% and pH 7.}
\end{figure}

 To investigate the nature of the
motion and to obtain information on different lengthscales we have
studied the momentum transfer $Q$ dependence of the relaxation time and
stretching parameter. In Fig.~\ref{fig:g(t)Q} the normalized
intensity correlation functions collected at different scattering
angles for a sample at $C_w$=0.20 \%, T=303 K and pH 7 are
reported. The behaviors of relaxation time and stretching
parameter as obtained through fits of the autocorrelation curves
according to Eq.~\ref{Eqfit} are shown as a function of the wave
vector $Q$ in Fig.~\ref{fig:tau beta vs Q}.

The relaxation time, reported in a double logarithmic plot, is
strongly $Q$ dependent, with a typical power law decay described
by the relation:
\begin{equation}
\tau=AQ^{-n} \label{EqPowerLaw}
\end{equation}
where $A$ is a constant and the exponent $n$ defines the nature of
the motion. In Fig.~\ref{fig:tau beta vs Q}a the fits according to
Eq.~\ref{EqPowerLaw} (full lines) are superimposed to the data
(symbols) and values of $n>$2 (in particular $n$ between 2 and 3)
are found, in agreement with those reported in previous studies on the same microgel ~\cite{MattssonNature2009} and on different
polymers~\cite{ColmeneroPRB1991, ColmeneroPRL1992}.

In Fig.~\ref{fig:tau beta vs Q}b the stretching parameter $\beta$
is reported, clearly showing no dependence on the wave vector $Q$.

\section{Conclusions}
\label{Conclusions} The swelling behavior of the PNIPAM-PAAc IPN microgel has been studied as a
function of temperature, pH, concentration and momentum transfer. The
presence of PNIPAM within the network determines the temperature
sensitivity behavior and the addition of PAAc introduces an
additional pH-sensitivity leading to interesting differences in
the transition process at acid and neutral pH, respectively. In particular we have found that the dynamical transition evidenced by looking at the relaxation time is more pronounced at pH 7 with respect to pH 5 reflecting a discontinuous and a continuous volume phase transition respectively.
Furthermore, while at neutral pH the
relaxation time reaches the highest value
at the highest concentration, at acid pH the dependence is inverted.
Finally at neutral pH the relaxation time for the highest
concentrated microgel shows, at odds with the low concentration
samples, an intriguing increase after the VPT transition. This
could be a precursor of the more complex behavior expected at even
higher concentrations where a thermoreversible gelation should be
observed~\cite{MattssonNature2009}. It would be very interesting
to address how the swelling capability and the gelation depend on
PAAc concentration.

\section*{Aknowledgments}
\label{Aknowledgments}R.A. acknowledge support from MIUR-PRIN.

\newpage

\section*{References}

%\bibliography{bib}
\bibliographystyle{unsrt}

\end{document}